\documentclass[journal,10pt]{IEEEtran}

\usepackage{amsmath,amssymb,amsfonts,amsthm}
\usepackage{bm}
\usepackage{cite}
\usepackage{url}
\usepackage{enumitem}
\usepackage[hidelinks]{hyperref}
\usepackage{graphicx}
\usepackage{bbm}
\newcommand{\ind}{\mathbbm{1}}
\usepackage{orcidlink}
\usepackage{setspace}
\usepackage{comment}

\usepackage{tikz}
\usetikzlibrary{arrows.meta,positioning,fit,calc}

\newtheorem{theorem}{Theorem}
\newtheorem{proposition}{Proposition}
\newtheorem{corollary}{Corollary}
\theoremstyle{definition}
\newtheorem{remark}{Remark}

\newcommand{\E}{\mathbb{E}}

\newcommand{\KL}{D}
\newcommand{\bits}{\{0,1\}}
\newcommand{\Bern}{\mathrm{Bern}}
\newcommand{\Inf}{\mathrm{Inf}}
\newcommand{\TotInf}{\mathrm{I}}

\title{Rule-Sensitive Sensor Participation in Distributed Detection with Prescribed Boolean Fusion Rules}

\author{
Ahmad Tanha~\orcidlink{0000-0003-3227-8667},~\IEEEmembership{Member,~IEEE}
\thanks{Ahmad Tanha is with the Signals, Telecommunications \& AI Department at Institut Polytechnique des Sciences Avancées (IPSA), Ivry-sur-Seine, France.
Email: \texttt{ahmad.tanha@ipsa.fr}.}
}

\begin{document}
\maketitle

\bstctlcite{IEEEexample:BSTcontrol}

\begin{abstract}
We study distributed binary detection when the fusion node must
implement a prescribed Boolean decision rule. Under conditionally
independent Bernoulli observations, we derive a signed-influence path
identity valid for arbitrary Boolean rules; for monotone rules, it
reduces to an exact Russo--Margulis representation in terms of
coordinate influences. The identity yields rule-sensitive necessary
conditions for a target Bayes error and an influence-weighted
Kullback--Leibler feasibility bound. We then connect Boolean relevance
to exact rule implementation and derive lower bounds on the number of
sensors that must participate, including a heterogeneous bound based
on the ordered per-sensor hypothesis separations. Numerical and Monte
Carlo results illustrate both the tightening provided by the
heterogeneous participation bound and the performance penalty that can
arise when a prescribed fusion rule is retained instead of optimal
likelihood-ratio fusion.
\end{abstract}


\section{Introduction}
\label{sec:introduction}

Distributed binary detection is a classical model for decentralized
inference in sensor networks and statistical signal processing.
Optimal fusion of binary sensor decisions was studied in
\cite{ChairVarshney1986}, decentralized detection under communication
constraints in \cite{ChamberlandVeeravalli2003}, and
communication-constrained hypothesis testing from an
information-theoretic perspective in \cite{AhlswedeCsiszar1986}.


In hard-decision distributed detection systems, the fusion rule may
be prescribed by legacy hardware, standardization requirements,
or constraints on implementation complexity and interpretability.
For example, counting rules have been studied for target detection
in wireless sensor networks~\cite{goel2018robustness}, while
\(k\)-out-of-\(N\) fusion has been considered in cooperative spectrum
sensing~\cite{maleki2011optimal}. Such prescribed rules need not
coincide with the statistically optimal fusion rule. This motivates
determining whether the available sensor statistics can support a
target error probability under a fixed Boolean rule and how many
sensors are indispensable for its exact implementation.

This problem differs from classical sensor selection. The methods in
\cite{joshi2009sensor,bajovic2011sensor} optimize the participating
subset for inference. Here, both the Boolean rule \(f\) and its
relevant-variable set \(\mathcal{S}_f\) are fixed; the sensors in
\(\mathcal{S}_f\) cannot generally be replaced by statistically
stronger ones. We instead derive statistical and cardinality
conditions that this prescribed set must satisfy.

Reporting-channel effects have been studied for binary-symmetric
multi-hop reporting~\cite{zhang2022decision}, \(M\)-ary relay
fusion~\cite{zhang2024mary}, and multi-bit multi-relay
fusion~\cite{zhang2025multibit}. Unlike these channel-aware fusion
designs, we analyze statistical feasibility and participation for a
prescribed Boolean rule under reliable one-bit reporting, then
specialize to binary symmetric channel (BSC) reporting errors.


We model the prescribed decision as a Boolean rule of form \(f:\bits^N\to\bits\)
and characterize it using signed and ordinary coordinate
influences~\cite{ODonnell2014}. Our contributions establish
rule-sensitive connections among Boolean influence, statistical
hypothesis separation, and exact sensor participation.

Our main contributions are summarized as follows.

\begin{itemize}[leftmargin=*,nosep]
    \item We derive a signed-influence path identity for arbitrary
    Boolean rules under heterogeneous product Bernoulli observations.
    For monotone rules, it becomes an exact Russo--Margulis
    coordinate-influence representation.

    \item We obtain rule-sensitive Bayes-error and influence-weighted
    Kullback-Leibler (KL) feasibility conditions linking sensor quality to the influence
    structure of the prescribed rule.

    \item We characterize the minimum participation required for exact
    rule implementation and derive a heterogeneous lower bound based
    on ordered sensor separations, together with a simpler
    maximum-separation relaxation.
\end{itemize}


\textbf{Notation.}
We write \([N]\triangleq\{1,\ldots,N\}\), and
\(\bm x_{\setminus i}\) denotes all coordinates of
\(\bm x\) except \(x_i\). The expression
\(f(\bm x_{\setminus i},b)\) evaluates \(f\) with
coordinate \(i\) set to \(b\in\bits\).
For \(\bm p=(p_1,\ldots,p_N)\in(0,1)^N\), let
\(\mu_{\bm p}\triangleq\bigotimes_{j=1}^N\Bern(p_j)\)
and
\(\mu_{\bm p,\setminus i}\triangleq
\bigotimes_{j\neq i}\Bern(p_j)\).
We use \(\ind\{\cdot\}\), \(\Bern(p)\),
\(\Pr_\nu\), \(\E_\nu\), \(D(P\|Q)\), and
\(h_2(\cdot)\) for the indicator, Bernoulli
distribution, probability, and expectation under
\(\nu\), KL divergence, and binary entropy,
respectively.

\section{System Model and Boolean Influence}
\label{sec:model}

We consider one-shot distributed binary detection with \(N\) sensors
and equal-prior hypothesis \(H\in\bits\). Sensor \(i\in[N]\) observes
\(X_i\in\bits\), where
\begin{equation}
X_i\mid H=h\sim\Bern(p_{h,i}),\qquad h\in\bits.
\label{eq:bernoulli-model}
\end{equation}


Here, \(X_i\) may represent either an inherently binary observation or
the hard-decision output of a fixed one-bit local quantizer; the design
of this quantizer is outside the scope of this work.

Under conditional independence~\cite{ChairVarshney1986}, for the binary vector
\(\bm X=(X_1,\ldots,X_N)\),
\begin{equation}
\begin{aligned}
P_h(\bm x)
&\triangleq \Pr(\bm X=\bm x\mid H=h)\\
&=\prod_{i=1}^{N}p_{h,i}^{x_i}(1-p_{h,i})^{1-x_i},
\quad \bm x\in\bits^N.
\end{aligned}
\label{eq:product-pmf}
\end{equation}

Let \(\bm p_h=(p_{h,1},\ldots,p_{h,N})\). Throughout the
aligned-sensing results, we assume
\begin{equation}
0<p_{0,i}\leq p_{1,i}<1,\qquad i\in[N],
\label{eq:aligned-sensing}
\end{equation}
so that \(X_i=1\) is at least as likely under \(H=1\).

The prescribed rule is a fixed Boolean mapping
\(f:\bits^N\to\bits\), and the fusion node must reproduce
\(\widehat H=f(\bm X)\); neither \(f\) nor the local observations are
design variables. Let \(\mathcal A\subseteq[N]\) be the participating
set. Each \(i\in\mathcal A\) reports \(X_i\) losslessly using one bit,
while sensors outside \(\mathcal A\) remain silent. Thus, the
participation cost is \(K\triangleq|\mathcal A|\), equal to the number
of transmitted bits in this model.

Define the rule-induced detection gap
\begin{equation}
\Delta_f
\triangleq
\E_{P_1}[f(\bm X)]-\E_{P_0}[f(\bm X)].
\label{eq:rule-gap}
\end{equation}
Since \(f\) is Boolean, its equal-prior error probability is
\begin{equation}
\begin{aligned}
P_e(f)
&=\tfrac12P_0(f(\bm X)=1)
+\tfrac12P_1(f(\bm X)=0)\\
&=\tfrac12(1-\Delta_f).
\end{aligned}
\label{eq:error-gap}
\end{equation}
Consequently, \(P_e(f)\leq\epsilon\) requires
\begin{equation}
\Delta_f\geq1-2\epsilon.
\label{eq:gap-condition}
\end{equation}
Although \(\Delta_f\) equals Youden's \(J\) statistic
\cite{Youden1950}, it is used here because it exactly determines the
error of the prescribed rule and admits the influence representation
below. In contrast, \(D(P_1\Vert P_0)\) depends only on the observation
laws, not on \(f\); Corollary~\ref{cor:kl-rule} connects it to the
rule-dependent gap.


Under the product measure \(\mu_{\bm p}\), define
the coordinate influence and signed coordinate
influence~\cite[Chs.~2--3]{ODonnell2014}
\begin{align}
\Inf_i^{\bm p}(f)
&\triangleq
\Pr_{\mu_{\bm p,\setminus i}}
\!\left[f(\bm X_{\setminus i},1)
\neq f(\bm X_{\setminus i},0)\right],
\label{eq:influence}\\
J_i^{\bm p}(f)
&\triangleq
\E_{\mu_{\bm p,\setminus i}}
\!\left[f(\bm X_{\setminus i},1)
-f(\bm X_{\setminus i},0)\right].
\label{eq:signed-influence}
\end{align}
They satisfy
\(\lvert J_i^{\bm p}(f)\rvert\leq\Inf_i^{\bm p}(f)\), and the total
influence is
\[
\TotInf_{\bm p}(f)\triangleq
\sum_{i=1}^N\Inf_i^{\bm p}(f).
\]
Direct differentiation gives
\begin{equation}
\frac{\partial}{\partial p_i}\E_{\mu_{\bm p}}[f]
=J_i^{\bm p}(f).
\label{eq:signed-derivative}
\end{equation}
If \(f\) is monotone increasing, then
\(J_i^{\bm p}(f)=\Inf_i^{\bm p}(f)\). Under the homogeneous measure
\(\bm p=(p,\ldots,p)\), this gives the classical Russo--Margulis
identity~\cite{ODonnell2014}
\begin{equation}
\frac{d}{dp}\E_{\mu_p}[f]
=
\sum_{i=1}^N\Inf_i^p(f)
=
\TotInf_p(f).
\label{eq:russo-margulis}
\end{equation}
Theorem~\ref{thm:path-identity} extends this representation to
heterogeneous parameters and arbitrary, possibly non-monotone, rules.

\textbf{Scope.}
The product-Bernoulli assumption enables an explicit influence
representation; correlated observations require a different
framework. Equal priors are needed only for
\eqref{eq:error-gap}, whereas the path identity is prior-independent.
Binary reporting isolates participation from quantizer design, and
independent reporting errors are treated in
Remark~\ref{rem:bsc} and Section~\ref{sec:examples_comparison}.
Thus, the influence profile governs statistical feasibility, while
the relevant-variable set governs exact implementation.

\section{Rule-Sensitive Detection Lower Bound}
\label{sec:detection_converse}

Define $\bm p_\theta=(1-\theta)\bm p_0+\theta\bm p_1,
\ \theta\in[0,1]$.
We first express the rule-induced detection gap exactly in terms of
the signed influences accumulated along the path between the two
hypothesis distributions as follows.

\begin{theorem}[Signed-influence path identity]
\label{thm:path-identity}
For any \(f:\bits^N\to\bits\),
\begin{equation}
\Delta_f
=
\sum_{i=1}^N(p_{1,i}-p_{0,i})
\int_0^1J_i^{\bm p_\theta}(f)\,d\theta.
\label{eq:path-identity}
\end{equation}
For monotone \(f\), \(J_i^{\bm p_\theta}(f)
=\Inf_i^{\bm p_\theta}(f)\), so \eqref{eq:path-identity} is an
exact Russo--Margulis coordinate-influence representation.
\end{theorem}

\emph{Proof idea:}
Differentiate \(\E_{\mu_{\bm p_\theta}}[f]\) along
\(\bm p_\theta\) and integrate the resulting coordinatewise
derivatives. For monotone rules, each signed effect equals the
corresponding influence.
The complete proof is given in
Appendix~\ref{app:proof_thm}.

Define the integrated signed and ordinary influences
\begin{align}
B_i(f)&\triangleq
\int_0^1J_i^{\bm p_\theta}(f)\,d\theta,
&
A_i(f)&\triangleq
\int_0^1\Inf_i^{\bm p_\theta}(f)\,d\theta.
\label{eq:integ1}
\end{align}
Then
\begin{equation}
\Delta_f
=
\sum_{i=1}^N(p_{1,i}-p_{0,i})B_i(f),
\qquad
|B_i(f)|\leq A_i(f),
\label{eq:integ2}
\end{equation}
with \(B_i(f)=A_i(f)\) for monotone rules.

This identity next yields a rule-sensitive necessary condition for
achieving a prescribed target error, presented below.

\begin{corollary}
\label{cor:target-error}
Under \eqref{eq:aligned-sensing}, if \(P_e(f)\leq\epsilon\), then
\begin{equation}
\sum_{i=1}^N(p_{1,i}-p_{0,i})A_i(f)
\geq1-2\epsilon.
\label{eq:necessary-integrated-influence}
\end{equation}
Consequently,
\begin{equation}
\|\bm p_1-\bm p_0\|_\infty
\int_0^1\TotInf_{\bm p_\theta}(f)\,d\theta
\geq1-2\epsilon.
\label{eq:necessary-total-influence}
\end{equation}
\end{corollary}

The result follows from \(\Delta_f\geq1-2\epsilon\),
\eqref{eq:integ2}, and \(|B_i(f)|\leq A_i(f)\). It gives a
rule-sensitive feasibility test: sensors with large separation
\(\delta_i=p_{1,i}-p_{0,i}\) contribute most effectively when they
also have large integrated influence \(A_i(f)\). Reliable detection
therefore requires alignment between sensor quality and rule
sensitivity. The complete proof is provided in
Appendix~\ref{app:proof_cor1}.

In the following, we further translate this error criterion into an
influence-weighted condition on the KL separation between the two
hypotheses.
\begin{corollary}
\label{cor:kl-rule}
Under \eqref{eq:aligned-sensing}, suppose \(P_e(f)\leq\epsilon\) and
define
\[
v_i\triangleq
\max_{\theta\in[0,1]}p_{\theta,i}(1-p_{\theta,i}).
\]
Then
\begin{equation}
D(P_1\Vert P_0)
\geq
\frac{(1-2\epsilon)^2}
{2\sum_{i=1}^N v_iA_i(f)^2}
\geq
\frac{2(1-2\epsilon)^2}
{\sum_{i=1}^N A_i(f)^2},
\label{eq:kl-rule}
\end{equation}
where the second inequality uses \(v_i\leq1/4\).
\end{corollary}

The proof combines a curvature lower bound on Bernoulli KL
divergence with weighted Cauchy--Schwarz. Thus,
Corollary~\ref{cor:kl-rule} complements the direct
error--influence relation by showing that the KL separation required
for a target error depends on the influence structure of the
prescribed rule. The complete proof is provided in
Appendix~\ref{app:proof_cor2}.

Corollaries~\ref{cor:target-error} and~\ref{cor:kl-rule} apply to
arbitrary Boolean rules; monotonicity is required only to replace
signed influences by ordinary influences with equality. 

We next
translate statistical feasibility into sensor participation
requirements for exact rule implementation.

\section{Sensor Participation Lower Bounds}
\label{sec:communication_lower_bounds}

Define the relevant-sensor set
\[
\mathcal S_f
\triangleq
\left\{
i\in[N]:
\exists\,\bm x_{\setminus i}\in\bits^{N-1}\, | \,
f(\bm x_{\setminus i},1)\neq f(\bm x_{\setminus i},0)
\right\}.
\]
Full support gives \(A_i(f)>0\iff i\in\mathcal S_f\).


We now characterize the structural participation required to
implement the prescribed Boolean rule exactly.
\begin{proposition}
\label{prop:exact-implementation}
Under full support, if \(\mathcal A\) permits exact implementation of
\(f\), then
\begin{equation}
\begin{aligned}
\mathcal S_f&\subseteq\mathcal A,\
K=|\mathcal A|\geq|\mathcal S_f|,\
K_{\min}(f)=|\mathcal S_f|.
\end{aligned}
\label{eq:exact-lb}
\end{equation}
\end{proposition}

The necessity follows by indistinguishability: omitting a relevant
sensor makes two inputs requiring different outputs generate identical
reports. Conversely, the reports from all relevant sensors determine
\(f\). This requirement concerns exact implementation of the prescribed
rule; it need not hold if the fusion node may replace \(f\) by another
rule with equal or smaller error. The complete proof, including sufficiency, is provided in
Appendix~\ref{app:proof_prop1}.

Let $\delta_i\triangleq p_{1,i}-p_{0,i},\
\delta_{(1)}\geq\cdots\geq\delta_{(N)},\
\delta_{\max}\triangleq\delta_{(1)}$.
For any \(m\), \(\sum_{j=1}^m\delta_{(j)}\) is the largest aggregate
separation attainable by an \(m\)-sensor set. It therefore provides a
best-case cardinality benchmark for the fixed set \(\mathcal S_f\),
not a rule for replacing its sensors. Provided that
\(\sum_{i=1}^N\delta_i\geq1-2\epsilon\), define
\begin{equation}
m^\star(\epsilon)
\triangleq
\min\left\{
m\in[N]:
\sum_{j=1}^{m}\delta_{(j)}\geq1-2\epsilon
\right\}.
\label{eq:mstar}
\end{equation}

Combining statistical feasibility with exact implementation yields
a heterogeneous lower bound on the number of participating sensors, addressed below.
\begin{corollary}
\label{cor:active-rate}
Under \eqref{eq:aligned-sensing}, let \(0\leq\epsilon<1/2\). If
\(P_e(f)\leq\epsilon\), then its fixed relevant set satisfies
\begin{equation}
\sum_{i\in\mathcal S_f}\delta_i
\geq1-2\epsilon.
\label{eq:fixed-set-separation}
\end{equation}
Under exact implementation,
\begin{equation}
K
\geq|\mathcal S_f|
\geq m^\star(\epsilon)
\geq
\left\lceil
\frac{1-2\epsilon}{\delta_{\max}}
\right\rceil.
\label{eq:active-lower-bound}
\end{equation}
If \(\sum_{i=1}^N\delta_i<1-2\epsilon\), no Boolean rule satisfying
\eqref{eq:aligned-sensing} can achieve \(P_e(f)\leq\epsilon\).
\end{corollary}

The ordered sum is used only to lower-bound the cardinality of the
fixed set \(\mathcal S_f\). Hence, \(m^\star(\epsilon)\) is a universal
best-case bound, and the prescribed rule may require more sensors when
its relevant coordinates are not the statistically strongest ones. The complete proof is provided in
Appendix~\ref{app:proof_cor3}.

\begin{remark}[Noisy reporting links]
\label{rem:bsc}
Under independent BSC reporting,
\(Y_i=X_i\oplus Z_i\), where \(Z_i\sim\Bern(q_i)\) and
\(0\leq q_i<1/2\),
\[
\begin{aligned}
Y_i\mid H&=h\sim\Bern(\widetilde p_{h,i}),\\
\widetilde p_{h,i}&=q_i+(1-2q_i)p_{h,i},\
\widetilde\delta_i=(1-2q_i)\delta_i.
\end{aligned}
\]
Thus, when the prescribed rule is applied to the received reports
\(f(\bm Y)\), the preceding results hold with \(p_{h,i}\) replaced by
\(\widetilde p_{h,i}\). Exact reproduction of \(f(\bm X)\) over noisy
links would require a separate reliable communication mechanism and is outside the scope of the one-shot model.
\end{remark}

\begin{figure*}[t]
\centering
\begin{minipage}[t]{0.315\textwidth}
    \centering
    \includegraphics[width=\linewidth]
    {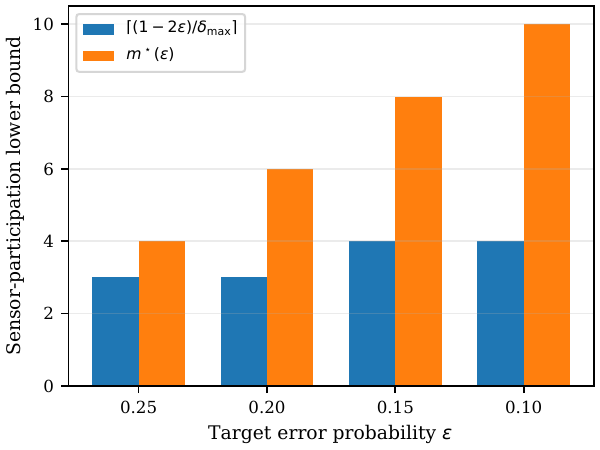}\\[-1mm]
    \textnormal{(a)}
\end{minipage}
\hfill
\begin{minipage}[t]{0.315\textwidth}
    \centering
    \includegraphics[width=\linewidth]
    {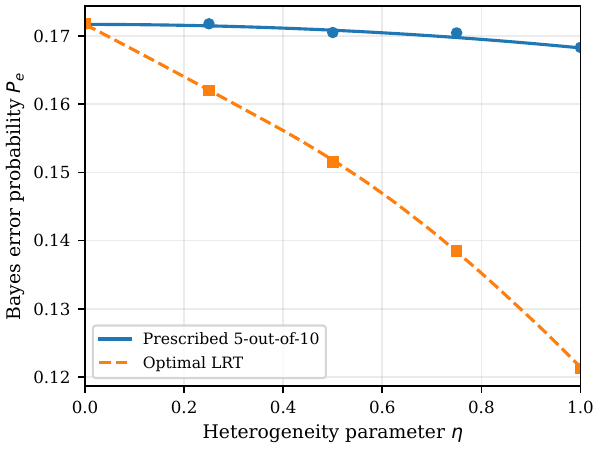}\\[-1mm]
    \textnormal{(b)}
\end{minipage}
\hfill
\begin{minipage}[t]{0.315\textwidth}
    \centering
    \includegraphics[width=\linewidth]
    {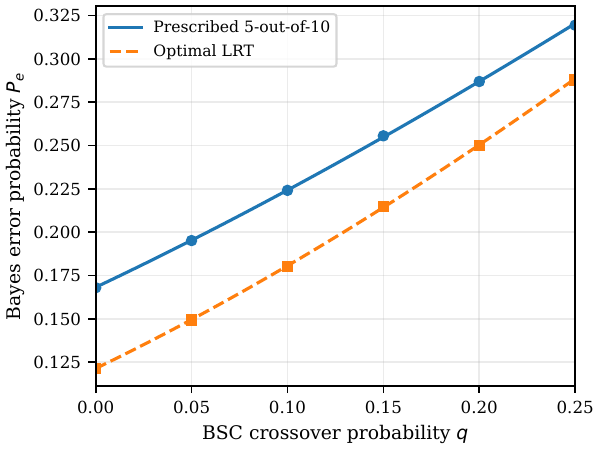}\\[-1mm]
    \textnormal{(c)}
\end{minipage}
\caption{Numerical results. (a) Heterogeneous
participation bound and its maximum-separation
relaxation. Bayes error of the prescribed
$5$-out-of-$10$ rule and optimal likelihood-ratio
test (LRT) versus (b) sensor heterogeneity at
$\bar\delta=0.30$ and (c) common BSC crossover
probability at $\eta=1$.}
\label{fig:numerical-results}
\end{figure*}


\begin{figure*}[t]
\centering
\begin{minipage}[t]{0.42\textwidth}
    \centering
    \includegraphics[width=\linewidth]
    {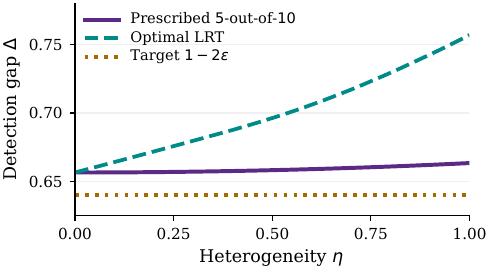}\\[-1mm]
    \textnormal{(a)}
\end{minipage}
\hfill
\begin{minipage}[t]{0.42\textwidth}
    \centering
    \includegraphics[width=\linewidth]
    {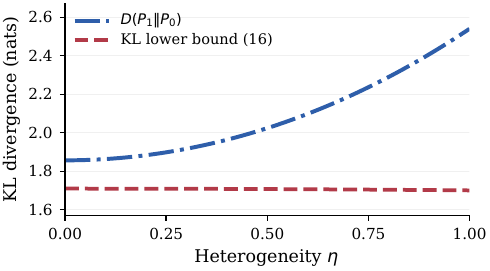}\\[-1mm]
    \textnormal{(b)}
\end{minipage}
\caption{Detection-gap and KL comparison under
heterogeneous Bernoulli sensing at $\epsilon=0.18$.
(a) Detection gaps of the prescribed
$5$-out-of-$10$ rule and optimal LRT, with the
target-feasibility threshold $1-2\epsilon$.
(b) Observation-level KL divergence and the
influence-weighted lower bound in
\eqref{eq:kl-rule}, evaluated for the prescribed
rule.}
\label{fig:gap-kl}
\end{figure*}


\section{Simulation Results}
\label{sec:examples_comparison}

For a non-monotone example, consider
\(f_{\oplus}(\bm x)=x_1\oplus x_2\). Although
\(\Inf_1^{\bm p}(f_{\oplus})
=\Inf_2^{\bm p}(f_{\oplus})=1\), its signed influences are
\(J_1^{\bm p}(f_{\oplus})=1-2p_2\) and
\(J_2^{\bm p}(f_{\oplus})=1-2p_1\). Thus,
Theorem~\ref{thm:path-identity} gives $\Delta_{f_{\oplus}}
=
\delta_1(1-p_{0,2}-p_{1,2})
+\delta_2(1-p_{0,1}-p_{1,1})$,
illustrating cancellations that ordinary influences do not capture.

For homogeneous sensing, \(p_{0,i}=p_0\), \(p_{1,i}=p_1\), and
\(\delta=p_1-p_0\), Corollary~\ref{cor:target-error} becomes
\[
\delta\int_0^1\TotInf_{\bm p_\theta}(f)\,d\theta
\geq1-2\epsilon.
\]
For the \(t\)-out-of-\(N\) rule,
\(f_t(\bm x)=\ind\{\sum_i x_i\geq t\}\),
\[
\Inf_i^{p}(f_t)
=
\binom{N-1}{t-1}p^{t-1}(1-p)^{N-t}.
\]
Every sensor is relevant for \(1\leq t\leq N\), so exact
implementation requires \(K=N\). The monotone specialization of
Theorem~\ref{thm:path-identity} also recovers the difference
\(\Pr(S_1\geq t)-\Pr(S_0\geq t)\), where
\(S_h\sim\operatorname{Bin}(N,p_h)\).

For comparison, the equal-prior Bayes-optimal LRT is
\begin{equation}
\begin{aligned}
f_{\mathrm{LRT}}(\bm x)
&=\ind\!\left\{\sum_{i=1}^N w_i x_i\geq\tau\right\},\\
w_i
&=\log\frac{p_{1,i}(1-p_{0,i})}
{p_{0,i}(1-p_{1,i})},\
\tau
=\sum_{i=1}^N\log\frac{1-p_{0,i}}{1-p_{1,i}}.
\end{aligned}
\label{eq:lrt}
\end{equation}
Unlike a prescribed counting rule, the LRT assigns unequal weights
to heterogeneous sensors. We measure the resulting penalty by
\(G_f\triangleq P_e(f)-P_e(f_{\mathrm{LRT}})\geq0\).
For homogeneous sensors, the weights are equal and the LRT reduces
to a counting rule with an appropriate threshold.

The rule-agnostic Fano bound gives \(K\geq1-h_2(\epsilon)\)
\cite[Chs.~2,~7]{CoverThomas2006}, whereas the exact-rule
requirement is \(K\geq m^\star(\epsilon)\geq
\lceil(1-2\epsilon)/\delta_{\max}\rceil\) by Corollary~\ref{cor:active-rate}.

\begin{table}[t]
\centering
\caption{Parameters and participation-bound results. Bound vectors
follow the ordering of the target-error vector.}
\label{tab:numerical_parameters}
\footnotesize
\setlength{\tabcolsep}{3pt}
\renewcommand{\arraystretch}{1.02}
\begin{tabular}{@{}ll@{}}
\hline
\textbf{Parameter} & \textbf{Value}\\
\hline
Number of sensors, \(N\) & \(10\)\\
\multicolumn{2}{c}{\emph{Participation-bound example}}\\
Sensor separations, \(\bm\delta\)
&
\(\begin{array}{l}
(0.20,0.12,0.10,0.08,0.07,\\[-1mm]
\phantom{(}0.06,0.05,0.05,0.04,0.03)
\end{array}\)\\
Target errors, \(\epsilon\)
& \((0.25,0.20,0.15,0.10)\)\\
\(m^\star(\epsilon)\)
& \((4,6,8,10)\)\\
Maximum-separation bound
& \((3,3,4,4)\)\\
\multicolumn{2}{c}{\emph{LRT and BSC experiments}}\\
Prescribed rule & \(5\)-out-of-\(10\)\\
Mean separation, \(\bar\delta\) & \(0.30\)\\
Heterogeneity, \(\eta\) & \([0,1]\)\\
BSC crossover probability, \(q\) & \([0,0.25]\)\\
Monte Carlo trials & \(10^6\) per point\\
\hline
\end{tabular}
\end{table}

For the participation example in Table~\ref{tab:numerical_parameters}
and Figure~\ref{fig:numerical-results}(a),
the heterogeneous bound is \((4,6,8,10)\), compared with
\((3,3,4,4)\) for the maximum-separation bound. The improvement grows
for stricter error targets because \(m^\star(\epsilon)\) accounts for
the ordered sensor qualities instead of treating every sensor as if
its separation were \(\delta_{\max}=0.20\).

For the LRT experiment, let
\[
\begin{aligned}
\delta_i(\eta)&=\bar\delta+\eta z_i,\\
p_{0,i}(\eta)&=\frac{1-\delta_i(\eta)}{2},\qquad
p_{1,i}(\eta)&=\frac{1+\delta_i(\eta)}{2},
\end{aligned}
\]
where \(\bar\delta=0.30\) and
\begin{align*}
    \bm z&=(0.25,0.20,0.15,0.10,0.05,\\
&-0.05,-0.10,-0.15,-0.20,-0.25)
\end{align*}

Thus, \(\eta\) varies sensor heterogeneity at fixed mean separation.
For Figure~\ref{fig:numerical-results}(b)--(c), exact errors enumerate
all \(2^{10}\) vectors; circles and squares show Monte Carlo estimates
from \(10^6\) equal-prior trials per operating point (base seed 2026),
using common observations for both rules. Solid/dashed curves are exact.

Figure~\ref{fig:numerical-results}(b) shows that the prescribed rule and
LRT coincide at \(\eta=0\), with \(P_e=0.1717\). At \(\eta=1\), their
errors are \(0.1683\) and \(0.1215\), respectively, giving a penalty
of \(0.0468\); the fixed counting rule cannot exploit heterogeneous
reliabilities as effectively as weighted fusion.
Figure~\ref{fig:numerical-results}(c) includes BSC reporting. At
\(q=0.10\), the corresponding errors increase to \(0.2243\) and
\(0.1807\). In both experiments, Monte Carlo estimates closely match
the exact probabilities.


Figure~\ref{fig:gap-kl} compares detection gaps and KL at
\(\epsilon=0.18\). In Figure~\ref{fig:gap-kl}(a), the prescribed-rule gap stays above
\(1-2\epsilon=0.64\), changing from \(0.6566\) to \(0.6635\) as
\(\eta\) grows from 0 to 1, whereas the LRT gap reaches \(0.7571\).
In Figure~\ref{fig:gap-kl}(b), \(D(P_1\Vert P_0)\) rises from \(1.8571\) to \(2.5397\),
while the Eq.~\eqref{eq:kl-rule} lower bound (using
\(A_i(f)\), \(v_i=1/4\)) changes from \(1.7103\) to \(1.7011\).
Unlike the rule-dependent gap, observation-level KL cannot
distinguish fusion rules applied to the same data.

\section{Conclusion}
\label{sec:conclusion}

We developed rule-sensitive error and KL feasibility conditions
and exact sensor-participation bounds for prescribed Boolean fusion.
The heterogeneous bound tightens its maximum-separation relaxation.
Numerical comparisons quantify the prescribed-rule penalty against
optimal LRT fusion, including BSC reporting. Correlated observations,
unequal priors, and general reporting channels remain open.

\nocite{tanha2026structured, tanha2024sensitivity, khalesi2026multiNLS, tanha2024influence, mirkarimi2026vbo, tanha2026low}

\clearpage
\section*{Appendices}
Appendix~\ref{app:proof_thm} proves Theorem~\ref{thm:path-identity}.
Appendices~\ref{app:proof_cor1} and~\ref{app:proof_cor2} prove
Corollaries~\ref{cor:target-error} and~\ref{cor:kl-rule}, respectively.
Appendix~\ref{app:proof_prop1} proves
Proposition~\ref{prop:exact-implementation}, and
Appendix~\ref{app:proof_cor3} proves Corollary~\ref{cor:active-rate}.
Appendix~\ref{app:numerical-reproducibility} describes numerical
reproducibility. Notation follows the definitions in the main text.

\appendices

\section{Proof of Theorem~\ref{thm:path-identity}}
\label{app:proof_thm}

Let \(F(\theta)=\E_{\mu_{\bm p_\theta}}[f]\). Since
\(F(1)-F(0)=\Delta_f\), the fundamental theorem of calculus, the
chain rule, and \eqref{eq:signed-derivative} give
\[
\Delta_f
=\int_0^1\sum_{i=1}^N
(p_{1,i}-p_{0,i})J_i^{\bm p_\theta}(f)\,d\theta,
\]
which is \eqref{eq:path-identity}. If \(f\) is monotone increasing,
\(f(\bm x_{\setminus i},1)-f(\bm x_{\setminus i},0)\in\{0,1\}\);
hence its expectation equals
\(\Inf_i^{\bm p}(f)\), so
\(J_i^{\bm p}(f)=\Inf_i^{\bm p}(f)\).

\section{Proof of Corollary~\ref{cor:target-error}}
\label{app:proof_cor1}

Let \(\delta_i=p_{1,i}-p_{0,i}\geq0\). From
\eqref{eq:gap-condition}, \eqref{eq:path-identity}, and
\(|B_i(f)|\leq A_i(f)\),
\[
1-2\epsilon
\leq\Delta_f
=\sum_i\delta_iB_i(f)
\leq\sum_i\delta_iA_i(f),
\]
proving \eqref{eq:necessary-integrated-influence}. Also,
\[
\sum_i\delta_iA_i(f)
\leq
\|\bm p_1-\bm p_0\|_\infty
\int_0^1\TotInf_{\bm p_\theta}(f)\,d\theta,
\]
which proves \eqref{eq:necessary-total-influence}.

\section{Proof of Corollary~\ref{cor:kl-rule}}
\label{app:proof_cor2}

For \(g_i(p)=\KL(\Bern(p)\Vert\Bern(p_{0,i}))\),
\(g_i(p_{0,i})=g_i'(p_{0,i})=0\) and
\(g_i''(p)=1/[p(1-p)]\geq1/v_i\) on
\([p_{0,i},p_{1,i}]\). Thus Taylor's theorem gives
\[
\KL(\Bern(p_{1,i})\Vert\Bern(p_{0,i}))
\geq\frac{\delta_i^2}{2v_i}.
\]
Since KL divergence is additive for product measures,
\(\KL(P_1\Vert P_0)\geq\sum_i\delta_i^2/(2v_i)\).
Therefore, by Cauchy--Schwarz,
\[
\Delta_f^2
\leq
\KL(P_1\Vert P_0)
\,2\sum_i v_iB_i(f)^2
\leq
\KL(P_1\Vert P_0)
\,2\!\sum_i v_iA_i(f)^2.
\]
Using \(\Delta_f\geq1-2\epsilon\) proves the first inequality in
\eqref{eq:kl-rule}; \(v_i\leq1/4\) gives the second.

\section{Proof of Proposition~\ref{prop:exact-implementation}}
\label{app:proof_prop1}

For \(i\in\mathcal S_f\), relevance yields some
\(\bm x_{\setminus i}\) such that
\(f(\bm x_{\setminus i},1)\neq f(\bm x_{\setminus i},0)\).
Full support makes both inputs possible. If
\(i\notin\mathcal A\), all participating sensors generate identical
reports for them, whereas exact implementation requires distinct
outputs. Hence
\(\mathcal S_f\subseteq\mathcal A\), so
\(K\geq|\mathcal S_f|\).

Conversely, if every sensor in \(\mathcal S_f\) reports, any two
inputs agreeing on \(\mathcal S_f\) differ only in irrelevant
coordinates. Changing such coordinates cannot change \(f\);
therefore \(\mathcal S_f\) determines \(f(\bm X)\). Activating
exactly these sensors is sufficient, and
\(K_{\min}(f)=|\mathcal S_f|\), proving \eqref{eq:exact-lb}.

\section{Proof of Corollary~\ref{cor:active-rate}}
\label{app:proof_cor3}

Since \(0\leq A_i(f)\leq1\) and, under full support,
\(A_i(f)>0\) iff \(i\in\mathcal S_f\),
\[
1-2\epsilon
\leq
\sum_{i\in\mathcal S_f}\delta_iA_i(f)
\leq
\sum_{i\in\mathcal S_f}\delta_i,
\]
which proves \eqref{eq:fixed-set-separation}. Let
\(m=|\mathcal S_f|\). Since the sum of the \(m\) largest
separations dominates that of any \(m\)-element set,
\(1-2\epsilon\leq\sum_{j=1}^{m}\delta_{(j)}\); hence
\(m\geq m^\star(\epsilon)\). Together with
Proposition~\ref{prop:exact-implementation},
\[
K\geq|\mathcal S_f|\geq m^\star(\epsilon).
\]
Moreover,
\[
1-2\epsilon
\leq
\sum_{j=1}^{m^\star(\epsilon)}\delta_{(j)}
\leq
m^\star(\epsilon)\delta_{\max},
\]
so
\[
m^\star(\epsilon)
\geq
\left\lceil
\frac{1-2\epsilon}{\delta_{\max}}
\right\rceil .
\]
Finally, if \(\sum_i\delta_i<1-2\epsilon\), then
\(\sum_i\delta_iA_i(f)\leq\sum_i\delta_i<1-2\epsilon\),
contradicting Corollary~\ref{cor:target-error}. This completes
\eqref{eq:active-lower-bound}.

\section{Information for Reproducibility}
\label{app:numerical-reproducibility}

The numerical results in Figures~\ref{fig:numerical-results}
and~\ref{fig:gap-kl} are obtained as follows.
For Figure~\ref{fig:numerical-results}(a), the
heterogeneous participation and maximum-separation bounds are evaluated
directly. For Figure~\ref{fig:numerical-results}(b)--(c), exact Bayes
errors are obtained by enumerating all \(2^{10}\) binary observation
vectors. Monte Carlo markers use \(10^6\) independent equal-prior
realizations per operating point, with base seed \(2026\); both fusion
rules are evaluated on the same samples.

Figure~\ref{fig:gap-kl} uses the same heterogeneous Bernoulli model,
with \(\eta\in[0,1]\) and target error \(\epsilon=0.18\). Panel (a)
evaluates the detection gaps of the prescribed \(5\)-out-of-\(10\) rule
and optimal LRT by exact enumeration, together with the feasibility
threshold \(1-2\epsilon=0.64\). Panel (b) computes
\(D(P_1\Vert P_0)\) and the influence-weighted lower bound in
\eqref{eq:kl-rule} using \(A_i(f)\) from 12-point Gauss--Legendre
quadrature. Here \(v_i=1/4\), so the two lower-bound expressions in
\eqref{eq:kl-rule} coincide. No Monte Carlo sampling is used for
Figure~\ref{fig:gap-kl}.

These evaluations use the parameters reported in
Section~\ref{sec:examples_comparison} and Table~\ref{tab:numerical_parameters}.

\clearpage
\bibliographystyle{IEEEtran}
\bibliography{refs}

\end{document}